\documentclass[
 amsmath,amssymb,aps,prb,twocolumn]{revtex4-1}

\usepackage{graphicx}
\usepackage{dcolumn}
\usepackage{bm}

\begin{document}

\preprint{APS/123-QED}

\title{Retrieving CeB$_6$'s lost magnetic entropy}

\author{Mehdi Amara}
\email{mehdi.amara@neel.cnrs.fr}
\author{Christine Opagiste}%
\author{Rose-Marie Gal\'era}%
\affiliation{Univ. Grenoble Alpes, CNRS, Grenoble INP\footnote{Institute of Engineering Univ. Grenoble Alpes.}, Institut N\'eel, 38000 Grenoble, France}


\date{\today}

\begin{abstract}
The reported temperature variations of CeB$_6$'s magnetic entropy are inconsistent with the fourfold degeneracy of the crystal field ground state. This old question is here addressed through new specific heat measurements and an improved description, in the cage context, of both the phonons and crystal field contributions to the specific heat. The antiferromagnetic transition is characterized as first-order and its latent heat determined. From the phonons' dispersion for a cage compound, the lattice specific heat contribution is derived from the LaB$_6$ data. Once corrected for the first-order transition and lattice contributions, the magnetic entropy displays the characteristic plateau of the quadruplet crystal field ground state, but at temperatures in excess of 30 K. Below 30 K, as the ordering temperature is approached, the magnetic entropy is substantially reduced. This anomalous temperature dependence is consistent with a crystal field ground state split by the rare-earth movement, a phenomenon specific to rare-earth cage compounds. 
\end{abstract}

\maketitle


\section{Introduction}
Since decades, the CeB$_6$ compound has been a center of interest in the fields of unconventional magnetism and heavy fermions physics. At low temperature and under zero magnetic field, it undergoes two orderings: from the paramagnetic state (phase I), an ordered state (phase II) develops at $T_Q$ = 3.3 K, then, at $T_N$ = 2.3 K, an antiferromagnetic state (phase III) is stabilized\cite{Kawakami1980, Takase1980, Effantin1985}. Between $T_N$ and $T_Q$, within phase II, the ordered state is reported to be non-magnetic and frequently interpreted as an antiferroquadrupolar order\cite{Hanzawa1984, Erkelens1987}, i.e. an order where 4$f$ electric quadrupoles alternate from site to site. However, this interpretation is difficult to reconcile with a number of observations\cite{Takigawa1983, Amara2012}. Beside the difficulties in the interpretation of the properties of this ordered state, the paramagnetic phase itself is not devoid of puzzles. There, the strong couplings of the 4$f$ electron with the conduction electrons and its lattice environment are already manifest. The measurements show a Kondo-like resistivity minimum around 150 K \cite{Takase1980, Ali1985} and a large contribution of the conduction electrons to the magnetic susceptibility, which result in a reduction of the apparent Ce$^{3+}$ magnetic moment \cite{Hacker1971, Kawakami1980}. Moreover, the strength of the cubic Crystalline Electric Field (CEF) is unusually large: inelastic neutron (INS) and Raman scatterings show that the $J = 5/2$ multiplet is split with a 540 K separation between the $\Gamma_7$ doublet and $\Gamma_8$ quadruplet \cite{Zirngiebl1984}. From the magnetic entropy variations, derived from specific heat measurements\cite{Fujita1980, PEYSSON1986}, the possibility of a $\Gamma_7$ doublet CEF ground state can be discarded.  Although the reported values at $T_Q$ = 3.3 K are close to the $R \ln 2$ J/(K mol) value of a doublet \cite{Lee1972, Fujita1980, PEYSSON1986}, the magnetic entropy steadily increases in the paramagnetic range, rapidly exceeding the doublet value. This variation is, however, hardly consistent with a quadruplet ground state: the quadruplet value is not reached below 40 K and no plateau, characteristic of a well isolated ground state, is to be observed. Also, the reported entropy value at $T_Q$, close to $R \ln 2$, is challenging the interpretation of phase II as a non-magnetic state. Indeed, according to the Kramers theorem applied to the Ce$^{3+}$ case, the minimal magnetic entropy within a non-magnetic state is precisely $R \ln 2$ J/(K mol). Starting from such an already reduced entropy, how is it then possible for the CeB$_6$ system to undergo, first, a non-magnetic transition and, second, an antiferromagnetic one? A better understanding of the physical mechanisms at play in CeB$_6$ requires to clarify this paramagnetic entropy issue. In the here introduced work, this question is addressed thanks to improvements along three directions: \newline
- the experimental determination of the specific heat, \newline
- the description of the phonons' contribution to the specific heat, \newline
- the theory, in order to account for the specific CEF effects in the cage context.\newline

\section{Experimental determination of the magnetic entropy}
\subsection{Specific heat measurements}
\label{SpecHeat}
In order to improve the experimental determination of the magnetic entropy of CeB$_6$, new specific heat measurements were performed. As regards the LaB$_6$, non-magnetic reference, single crystals grown in Kiev, by the team of Dr. N. Shitsevalova, were used. These crystals were obtained from borothermal reduction of La$_2$O$_3$, under vacuum at 1750 $^{\circ}$C, using amorphous natural boron. The obtained powder was successively pressed into rods and sintered at 1800 $^{\circ}$C. The rods were then processed by zone melting under argon atmosphere, resulting in large, single phase, LaB$_6$ crystals. The used CeB$_6$ crystals are from older batches, similarly processed in Sendai by Prof. S. Kunii. They were lent to us by Dr. L. P. Regnault. These high-quality single crystals, initially intended for neutron scattering experiments, were produced using $^{11}$B enriched boron, whereas natural boron was used for our LaB$_6$ reference. All the used crystals were received in form of oriented platelets. The small specific heat samples of masses $m =$ 6 mg for LaB$_6$ and $m =$ 2.95 mg for CeB$_6$, were subsequently cut from these.\newline
The specific heat measurements were performed using the relaxation technique in an automated Quantum Design PPMS system. Two cryogenic configurations were used, the normal $^4$He flux one, for temperatures between 1.8 K and 60 K, and the additional closed-cycle $^3$He insert for temperatures down to 0.6 K (in the case of CeB$_6$ only). The thermal coupling between the sample and the setup platform was improved by use of a very small quantity of Apiezon N grease, which is accounted for thanks to the addenda measurements.

\begin{figure}
\includegraphics[width=\columnwidth]{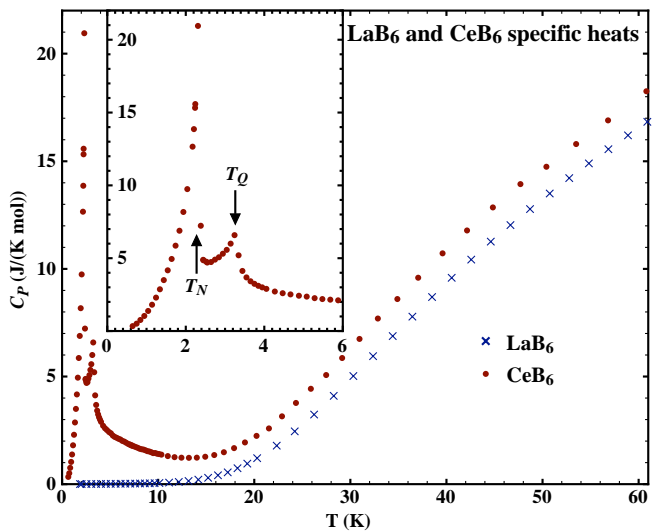}
\caption{\label{ChalSpe} The temperature dependence of the specific heat for LaB$_6$ (crosses) and CeB$_6$ (round dots), as determined using the relaxation technique. The inset gives the low temperature detail for CeB$_6$, showing the transitions at $T_N$ and $T_Q$. Note the very sharp anomaly at $T_N$.}
\end{figure}

\begin{figure}
\includegraphics[width=\columnwidth]{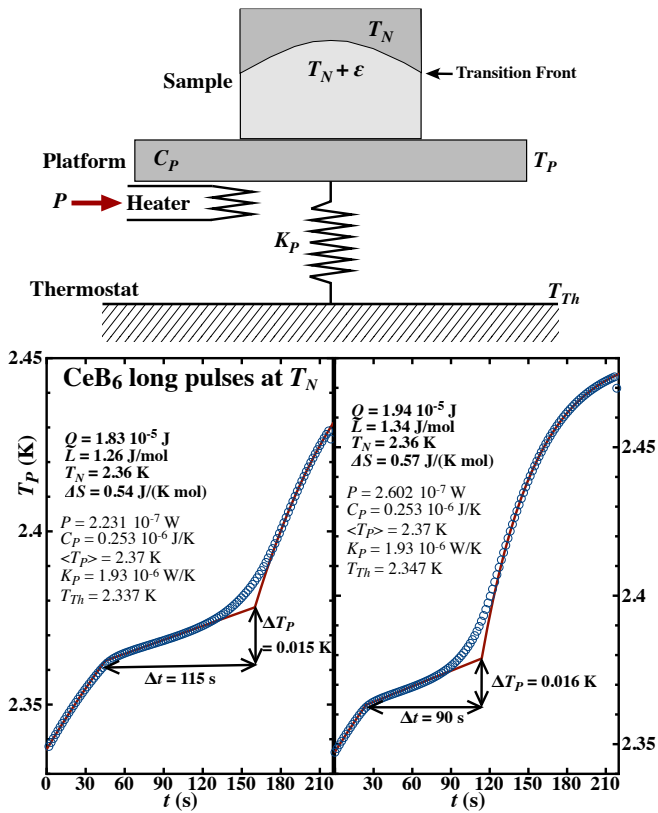}
\caption{\label{LongPulses} Upper part: Thermal exchange scheme during a first-order transition in heating conditions. The sample's platform has heat capacity $C_P$, thermal resistance to the thermostat $K_P$ and absorbs a heating power $P$. As the transition front moves, instead of being constant, the platform temperature $T_P$ increases. Lower part: Two examples of $T_P (t)$ temperature profiles, showing the crossing of the antiferromagnetic transition in CeB$_6$, when using long heating pulses. These processes have the same duration but differ in their starting temperature and temperature rise (related to the heating power $P$). The full lines are adjusted considering exponential relaxation laws above and below a linear "plateau" of duration $\Delta t$ and temperature rise $\Delta T_P$. }
\end{figure}

The LaB$_6$, non-magnetic reference curve (see Fig. \ref{ChalSpe}), was obtained using the default two time constants (two-tau) fitting of the relaxation process, as provided by the PPMS software. The same options were used for the CeB$_6$ sample, but this time using the $^3$He insert. Fig. \ref{ChalSpe} shows the resulting specific heat curve for CeB$_6$. Inspecting the collected data for CeB$_6$, it appeared that the two-tau fitting was systematically failing at the top of the antiferromagnetic anomaly at $T_N$. As shown on Fig. \ref{ChalSpe}, the processed relaxation data display a very sharp anomaly at $T_N$, in agreement with the literature. This aspect of the anomaly and the failure of the fitting process at $T_N$, made us suspect a first-order antiferromagnetic transition. To clarify this point, a second series of measurements on the same sample, using the standard $^4$He setup, was carried out, focusing on the transition at $T_N$. This time, the system was forced to proceed by using long heating pulses, with adjusted duration and temperature amplitude, which affect the heating power, in order to cover the whole transition process. The lower part of Fig. \ref{LongPulses} shows two examples of time dependence of the platform temperature, for long pulses processes crossing the antiferromagnetic transition. The inflection in the temperature evolution, that reflects the expected first-order plateau at $T_N$, is well evidenced. The "plateau" actually displays a significant slope, which can be understood considering that the sample cannot be of uniform temperature (see the thermal exchange diagram of Fig. \ref{LongPulses}, upper part): as the transition front moves though the sample, the platform temperature keeps increasing, but at a slower pace, until the "cold" point at $T_N$ disappears with the last fraction of antiferromagnetic CeB$_6$. The difficulty then lies in defining the duration $\Delta t$ of the transition and the corresponding temperature rise $\Delta T_P$ of the platform. This is done here by idealizing the temperature profile during the transition, replaced by a constant slope process. The curves before and after the "plateau" can be fitted with simple relaxation exponentials, with identical temperature limit, but different time constants due to different sample specific heat above and below the transition. The process is analyzed considering only conductive heat exchange (coupling constant $K_P$ between the platform and the thermostat), a stable thermostat at temperature $T_{Th}$ and constant heating power $P$. During the transition process, the heat absorbed by the sample is:
\begin{equation*}
Q = \left[ P-K_P (\langle T_P \rangle - T_{Th}) \right] \Delta t -C_P \Delta T_P
\end{equation*}
where $\langle T_P \rangle$ is the time averaged temperature of the platform during the process and $C_P$ the platform heat capacity (including the thermal contact grease). For a good coupling between the platform and sample, the same temperature rise $\Delta T_P$ will occur at the sample's face in contact with the platform. As the sample's fraction that hasn't undergone the transition maintains a $T_N$ temperature, there is a thermal gradient between the transition front and the platform (Fig. \ref{LongPulses}, upper part). At the end of the process, when the last fraction at $T_N$ disappears, the sample is, on average, overheating with respect to $T_N$. In the simplest case of a linear temperature profile and a regular sample's shape, the sample's average temperature is then close to $(T_P + T_N )/2$. This means that the extent of the overheating with respect to $T_N$ is $(T_P - T_N )/2 = \Delta T_P /2$. In the definition of the latent heat $L$ of the transition, one can account for this excess of heat transfer to the sample:
\begin{equation*}
L= Q -C_S (T_{N_+}) \frac{\Delta T_P} {2}
\end{equation*}
where $C_S (T_{N_+})$ is the sample specific heat immediately above the transition temperature. Using the thermal parameters of the PPMS "puck" (see Fig. \ref{LongPulses}), averaging the values derived from the two pulses and estimating the uncertainty on such a determination as not better than 10 \%, one obtains :
\begin{equation*}
L = 1.30 \pm 0.13\;  \text{J/mol}
\end{equation*}
The corresponding change in entropy for the transition at $T_N=$ 2.36 K is:
\begin{equation*}
\Delta S = L/T_N = 0.55 \pm 0.06\;  \text{J/(K mol)}
\end{equation*}
The characterization of the antiferromagnetic transition as first-order allows to recover some of the missing paramagnetic entropy of CeB$_6$. This correction represents about 5\% of the entropy of a quadruplet ground state.

\subsection{Phonons' contribution to the specific heat}
At this point, the remaining difficulty for extracting the magnetic part of the specific heat, then the magnetic entropy of CeB$_6$, is the proper identification of the non-magnetic contributions. This requires to determine the temperature dependence of the specific heat for a non-magnetic element in the series, in the present case LaB$_6$.
In a non-magnetic metal, the specific heat $C$ is usually well described by separating two contributions, one from the conduction electrons, $C_{e}$, the other from the phonons, $C_{ph}$:
\begin{equation}
C(T) = C_{e}(T) +C_{ph}(T)
\end{equation}
Due to the very low compressibility of solids, no distinction is here made between the constant pressure and the constant volume specific heat. At low temperature, where it is influent, the electronic term can be reduced to the linear form $C_{e}(T) = \gamma T$, where $\gamma$ is the specific heat electronic constant. As regards the phonon term $C_{ph}$, the most common approach is to describe it using the Debye approximation which, in the low temperature limit, yields the cubic temperature term. This term reflects the low frequency acoustic modes and, in principle, allows to derive, from a non-magnetic reference, the phonons contribution for a magnetic element in the series. In the simplest harmonic approach, the forces that determine the springs stiffnesses in a classical model are maintained, whereas the inertia increases across the series.  In this scheme, the eigenfrequencies, as well as the Debye temperature, scale via the square root of the formula masses ratio. If necessary, more than one Debye temperature are introduced \cite{Hofmann1956}.\newline
This Debye approach is known to fail in the description of cage compounds, where the low temperature dependence of the phonon specific heat cannot be reduced to a cubic term. In these systems, the weakly dispersive rattling of the cage guest yields contributions closer to the Einstein approach than to the dispersive (acoustic) Debye one. In the case of rare-earth hexaborides, there has been attempts\cite{Smith1985, Kunii1997, Mandrus2001} at describing the phonon contribution by using the Einstein model or an empirical mixture of Einstein and Debye. This increases the number of involved parameters and, by lack of an underlying physical model, it is difficult to scale them from one element to another in the series.

\paragraph{Phonons dispersion in a cage system}
\begin{figure}
\includegraphics[width=\columnwidth]{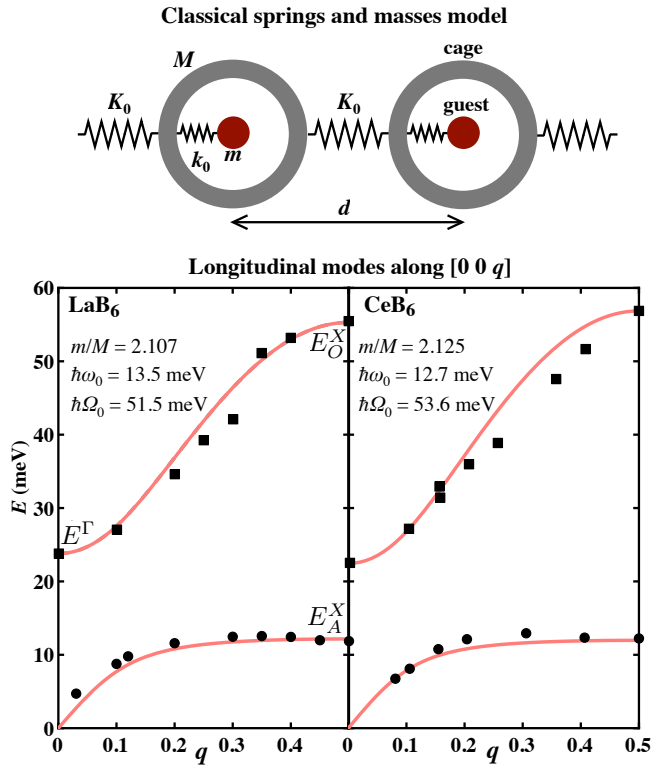}
\caption{\label{CageDisp} Upper part: the chain of masses ($m$ for the guest and $M$ for the cage) and springs (stiffnesses $k_0$,  $K_0$) used for the harmonic, classical description of lattice modes in a cage system. Lower part: the inelastic neutron scattering data for LaB$_6$ (left, from Ref. \onlinecite{Smith1985}) and CeB$_6$ (right, from Ref. \onlinecite{Kunii1997}), showing the dispersion of the longitudinal mode along the fourfold direction. Superimposed are the dispersions curves (full lines) derived from the above model for the indicated values of mass ratios and frequencies. }
\end{figure}

In light rare-earth hexaborides, the two lowest phonon dispersion branches are well reproduced by a harmonic model consisting in a chain, of period $d$, of identical rigid cages of masses $M$ (see the upper part of Fig. \ref{CageDisp}), interconnected by springs of stiffness $K_0$ (see Ref. \onlinecite{Amara2019}). In each cage a mass $m$, the rare-earth, is attached by a spring of stiffness $k_0$. Writing the classical equations of motion for small deviations from the equilibrium positions along the chain axis, introducing a propagating wave at frequency $\omega$ and wave vector $q$ (here in unit $2\pi /d$), one obtains the relation:
\begin{equation}
\label{EqDisp1}
\cos (2\pi q) = 1 - 2 (1 + \frac{m}{M} \frac{{{\omega _0}^2{\kern 1pt} }}{{{\omega _0}^2 - {\kern 1pt} {\omega ^2}}})\frac{{{ \omega ^2}}}{{{\Omega _0}^2}}
\end{equation}
were ${\omega_0}^2 = k_0  / m$ is the natural frequency of the "rattler" and  ${\Omega_0}^2 = 4 K_0 / M$ the top frequency for a chain of empty cages.
This allows to define the two branches of the dispersion curves, here written by introducing $x=\omega/ \omega_{0}$, the mass ratio $\alpha = m/M$, the frequency ratio $\rho = \Omega_{0}/ \omega_{0}$ and the function $K(q) = {\rho^2} (1- \cos(2\pi q))/2$:
\begin{equation}
\label{EqDisp2}
{\omega_{\pm}(q)} = \omega_{0} \sqrt {\frac{{1 + \alpha  + K(q) \pm \sqrt {{{(1 + \alpha + K(q))}^2} - 4 K(q)} }}{2}}
\end{equation}
where the $+$ and $-$ options respectively give the expression for the optical and acoustic branches.

\begin{figure}
\includegraphics[width=\columnwidth]{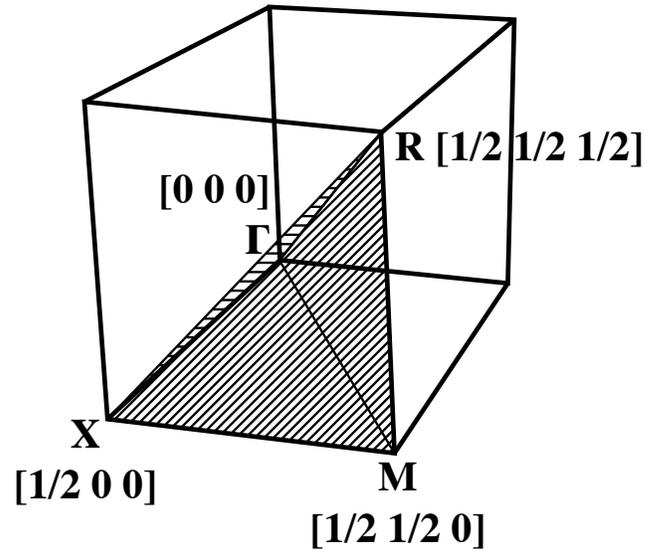}
\caption{\label{FirstZB} The positive octant of the cubic first Brillouin zone showing the characteristic points $\Gamma$, $X$, $M$ and $R$. These points are the vertices of a tetrahedron (hatched faces) that, under the cubic transformations, generates the whole first Brillouin zone. In the numerical calculation of the phonons' specific heat, all the considered samples belong to this representative volume. }
\end{figure}

Fig. \ref{CageDisp} shows the dispersion curves for longitudinal waves propagating along $[0 \; 0 \; q]$ in LaB$_6$ (from Ref. \onlinecite{Smith1985}) and CeB$_6$ (from Ref.  \onlinecite{Kunii1997}). These dispersion curves have three characteristic points, at positions $\Gamma$ and $X$ in the cubic first Brillouin zone (see Fig. \ref{FirstZB}), that allow a direct determination of the parameters $\omega_0$ and $\Omega_0$:\newline
- the $q=0$ point ($\Gamma$ point of the cubic zone), where converge all the optical branches at the energy $E^{\Gamma} = \hbar \omega_{+}(0) $. From Eq. (\ref{EqDisp1}), ${\omega_{+}(0)=\omega_0 \sqrt{1+\alpha}}$. Using the mass ratios $\alpha$ for 98\% enriched boron, one obtains : $\hbar \omega_0 = 13.5$ meV for LaB$_6$ and $\hbar \omega_0 = 12.7$ meV for CeB$_6$.\newline
- the zone border $q=1/2$ ($X$ point in the $[0 \; 0 \; q]$ direction), for the acoustic and optical branches, respectively yielding the energies $E^{X}_A$ and $E^{X}_O$. From  Eq. (\ref{EqDisp1}), one can identify the frequency $\Omega_0$ for these longitudinal waves:\newline
 ${\hbar ^2}{\kern 1pt} {\Omega _0}^2 = {\left( {E_A^X} \right)^2} + {\left( {E_O^X} \right)^2} - {\left( {{E^\Gamma }} \right)^2}$\newline
In this way, one obtains: $\hbar \Omega_0 = 51.5$ meV for LaB$_6$ and $\hbar \Omega_0  = 53.7$ meV for CeB$_6$.\newline
However, caution is required as regards the precision of these determinations. The used dispersion curves were obtained from inelastic neutron scattering on triple axis spectrometers. In such conditions, the error on an energy determination can exceed one percent. Using the above values, computed curves are superimposed on the experimental data in Fig. \ref{CageDisp}. The agreement is very satisfactory considering the above evoked uncertainties and the simplicity of the model, dependent on only two parameters.\newline

For small deviations from the equilibrium positions, the motion equations that yield Eq. (\ref{EqDisp1}) for a chain generalize to a three-dimensional lattice. The elements in the chain are replaced by infinite planes of guests or cages with parallel deviations. One only needs to consider the relevant $d$ spacing and to adapt the equations by considering appropriate stiffnesses for the particular directions $\widehat{\textbf{q}}$, of the wave vector, and $\widehat{\textbf{p}}$, of the polarization. The dispersion relations can be thus generalized to any directions of propagation and polarization. As regards the wave vector direction $\widehat{\textbf{q}}$, the associated chain period $d$ identifies with the smallest spacing between consecutive lattice planes perpendicular to $\widehat{\textbf{q}}$. This minimal spacing defines the smallest, physically relevant wavelength on the lattice, $\lambda = 2d$, which, in the reciprocal space, is associated with the first zone border vector $\mathbf{B_q} $, parallel with $\mathbf{q}$. On the segment from the origin to $\mathbf{B_q}$, the dispersion relations keep the forms of Eq. (\ref{EqDisp1}), provided one replaces $\cos(2\pi q)$ with $\cos (\pi \frac{|\mathbf{q}|}{|\mathbf{B}_{\mathbf{q}}| })$ and defines $K(q)$ as $K(q)=\rho^2 (1- \cos (
\pi \frac{|\mathbf{q}|}{|\mathbf{B}_{\mathbf{q}}| }))$ in Eq. (\ref{EqDisp2}). As the frequency $\omega_0$, and its associated $k_0$ stiffness, are isotropic in a cage with $O_h$ symmetry at its center, they apply for all wave vectors and polarizations. In the generalization of the dispersion relation, only the top frequency $\Omega_0$, for the lattice of empty cages, has to be adapted to the wave polarization $\widehat{\textbf{p}}$ and propagation direction $\widehat{\textbf{q}}$, replacing the constant $\Omega_0$ with the function $\Omega_0 (\widehat{\textbf{q}},\widehat{\textbf{p}})$ in Eq. (\ref{EqDisp1}). In search for a simplification, it is assumed that using a constant $\widetilde{\Omega}_{0}$, in place of the function $\Omega_0 (\widehat{\textbf{q}},\widehat{\textbf{p}})$, can result in a satisfactory description of the low temperature specific heat of a hexaboride. This is inspired by the Debye approximation, but provides a more realistic description of the cage context: the characteristic features of the dispersion curves, in particular the flattened acoustic branch surmounted by an energy gap, are preserved. In this way, only two parameters are required for the description of the phonons's contribution: the rattler frequency $\omega_0$ and $\widetilde{\Omega}_{0}$. The parameter $\widetilde{\Omega}_{0}$ represents the lattice of empty boron cages. As such, it should not vary much across the RB$_6$ series, except for the anharmonic effect of a slight reduction in the lattice parameter due to the lanthanide contraction. 

\paragraph{Cage system specific heat}
\begin{figure}
\includegraphics[width=\columnwidth]{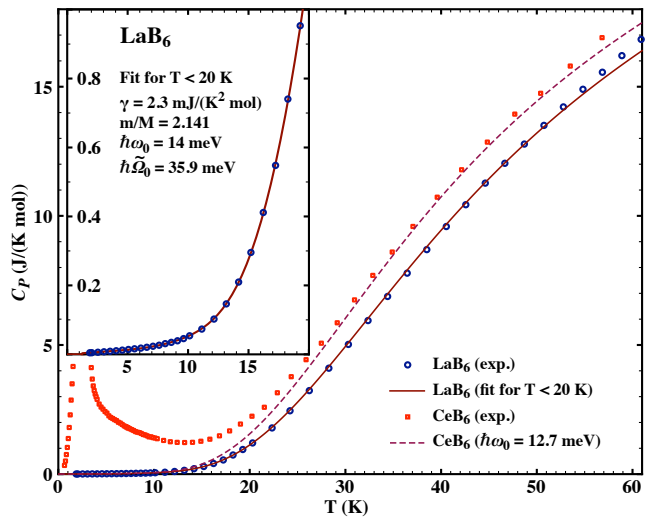}
\caption{\label{FitLaB6} The specific heat for LaB$_6$ (circles) and CeB$_6$ (squares) confronted with calculations based on the generalized dispersion curves and including the electronic term (defined by the $\gamma$ constant). The full lines show the computed curves for LaB$_6$, using $\omega_0$ and $\widetilde{\Omega}_{0}$ values deduced from a fit below 20 K (see detail and values in the inset). The dashed line is an attempt at describing the non-magnetic background of CeB$_6$ using the neutron data for $\omega_0$ and a $\widetilde{\Omega}_{0}$ value derived from the LaB$_6$ fit.}
\end{figure}

Thanks to the above simplification, for any $\mathbf{q}$ within the first Brillouin zone, one can associate two frequencies $\tilde \omega_{\pm} (\mathbf{q})$ for each of the three polarizations. Then, at temperature $T$, the phonons specific heat $C_{ph}$ can be computed by summing, within the first Brillouin zone, over all wave vectors and polarizations describing the vibration modes:

\begin{equation}
 C_{ph} (T) = 3\sum\limits_{\mathbf{q}} {k_{\text{B}}{{\left( {\frac{{\hbar \,\tilde \omega (\mathbf{q})}}{{k_{\text{B}}T}}} \right)}^2}\frac{{{e^{ - \frac{{\hbar \,\tilde \omega (\mathbf{q})}}{{k_{\text{B}}T}}}}}}{{{{\left( {1 - {e^{ - \frac{{\hbar \,\tilde \omega (\mathbf{q})}}{{k_{\text{B}}T}}}}} \right)}^2}}}} 
\end{equation}

In practice, the summation can be performed for a discrete fraction of the first Brillouin zone, considering, inside the positive octant, samples at a number $N$ of $\mathbf{q}$ nodes on a cubic lattice. In direct space, this amounts to the calculation of the specific heat for a crystal consisting in $N$ unit cells. Obviously, $N$ needs to be large enough to approach the macroscopic limit. This requirement can substantially slow down the calculation and it is preferable to take advantage of the cubic symmetry. Redundant contributions can be avoided by restricting to nodes included in a representative polyhedron as represented on Fig. \ref{FirstZB}: each considered $\mathbf{q}$ node accounts for its symmetry equivalents by considering a multiplicity factor. In Fig. \ref{FitLaB6}, this method of calculation is used for describing the LaB$_6$ experimental data. The displayed curves are computed considering 364x3 representative samples of the first Brillouin zone, equivalent to 1728x3 modes in the positive octant. At the graph's scale, calculations for as few as 56x3 representative samples are indistinguishable from the displayed curves. The electronic term is deduced from the linear, low temperature part of the curve. As regards the phonons contribution, only two parameters are active, $\omega_0$ and $\widetilde{\Omega}_{0}$, for adjusting the computed curve to the experimental data. The values appearing in Fig. \ref{FitLaB6} are obtained by a mean-square fit on the data for temperatures lower than 20 K (inset of  Fig. \ref{FitLaB6}). The calculated curves, based on the low temperature data, extrapolate very well up to 50 K. Above 50 K, the cost of the simplification of the dispersion curves, here adapted to the low temperature specific heat, starts to materialize: a better description in this temperature range would require to increase the  $\widetilde{\Omega}_{0}$ value. Looking at the refined values for $\omega_0$ (reported in Fig. \ref{FitLaB6}), $\hbar \omega_0 =$ 14 meV is larger than the inelastic neutron scattering determination at 13.5 meV, but the difference may be within the uncertainty of the neutron determination (which values are not explicited in Ref. \onlinecite{Smith1985, Kunii1997}).\newline
In order to adapt the phonon contribution from the LaB$_6$ reference, to the CeB$_6$ case, one can think of resorting to a mass scaling in the harmonic model context. This is done in the hypothesis that the underlying forces, resulting in the stiffnesses of the elastic description, are maintained in CeB$_6$. Then, one should correct the eigenfrequencies by a factor reflecting the ratios of the atomic masses, $\sqrt \frac{M_{La}}{M_{Ce}}$ for $\omega_0$, and $\sqrt \frac{M_{B_{nat.}}}{M_{B_{enr.}}}$ for $\widetilde{\Omega}_{0}$. The later correction is required in order to account for the use of natural boron ($B_{nat.}$) in the synthesis of LaB$_6$ as opposed to 98\% $^{11}$B enriched boron in the case of CeB$_6$ ($B_{enr.}$). One thus derives the following values for the phonons' contribution parameters in CeB$_6$:\newline
 $m/M = 2.125$, $\hbar\omega_0 = 13.9 \text{ meV}$ and $\hbar\widetilde{\Omega}_{0} = 35.6 \text{ meV}$.\newline
This mass scaling has a tiny effect, resulting in a curve almost undistinguishable from that of LaB$_6$ at the scale of Fig. \ref{FitLaB6}. It cannot account, even in presence of magnetic effects, for the specific heat difference observed above 20 K between the CeB$_6$ and LaB$_6$ curves of Fig. \ref{ChalSpe}. The failure of the mass scaling points to the limits of the harmonic approximation, despite its seemingly satisfactory description of the dispersions. In particular, the cage context implies larges amplitude excursion for the rare-earth within the rigid limits of the cage. This alone would determine an anharmonic potential, even in absence of magnetic effects, as it implies a sharply rising potential close to the boron framework. In the simplest picture, one can expect the smaller Ce$^{3+}$ ion to have more room than La$^{3+}$, which, independently of the mass correction, would result in a lower $\omega_0$ frequency for CeB$_6$. \newline
As regards the $\omega_0$ value for CeB$_6$, a more empirical option is the neutron spectroscopy determination of $\hbar \omega_0$ that yields a 12.7 meV. In the hypothesis that the harmonic correction can still be applied to the rigid boron framework, where small deviations from the equilibrium position are granted, the mass scaled value for $\widetilde{\Omega}_{0}$ is maintained. This defines a second set for the parameters defining the CeB$_6$ non-magnetic contribution: \newline
$m/M = 2.125$,  $\hbar\omega_0 = 12.7$  meV and  $\hbar\widetilde{\Omega}_{0} = 35.6$ meV.\newline
This time, the corresponding curve in Fig. \ref{FitLaB6} is much closer to the CeB$_6$ data (dashed line for $\hbar\omega_0 = 12.7 \text{ meV}$). This supports the idea that the large difference in the background specific heat, with respect to the LaB$_6$ reference, is essentially due to the frequency $\omega_0$ of the guest inside the cage. Indeed, in the temperature range of interest, the computed curve is very sensitive to the $\omega_0$ value. Considering the uncertainty on the neutron scattering data, and the 0.5 meV discrepancy observed in the LaB$_6$ case, the agreement of $\hbar\omega_0 = 12.7 \text{ meV}$ with the CeB$_6$ background is somewhat lucky. Note that the value used for the $\gamma$ electronic constant, of little influence above 20 K, is the same as the one derived from the LaB$_6$ low temperature data (inset of Fig. \ref{FitLaB6}).

\subsection{The magnetic entropy of CeB$_6$}
\begin{figure}
\includegraphics[width=\columnwidth]{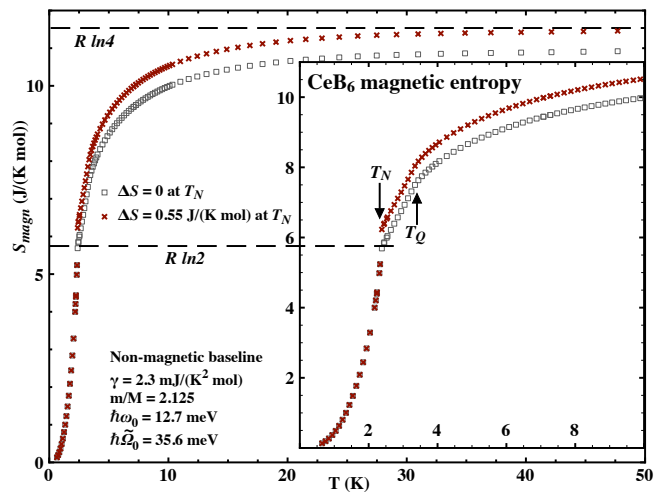}
\caption{\label{Entropy} Temperature variation of CeB$_6$ magnetic entropy. The cross and square symbols respectively show data corrected, or uncorrected, for the first-order transition at $T_N$. The inset gives the detail of this variation for T $ < $ 10 K. All curves are corrected for the non-magnetic contributions to the specific heat (computed using the displayed parameters). The dashed horizontal lines refer to the entropy values for a doublet ($R \ln2$) and a quadruplet ($R \ln4$). The latter is the expected value for an isolated $\Gamma _8$ crystal field ground state.}
\end{figure}

Despite the uncertainty on the $\omega_0$ value, the computed curve for CeB$_6$ in Fig. \ref{FitLaB6} provides a likely non-magnetic background. One can then proceed with the determination of the magnetic entropy of CeB$_6$ by subtracting this background from the specific heat in order to define the magnetic contribution $C_{mag}(T)$. Between zero temperature and 0.6 K, the missing CeB$_6$ specific heat data are interpolated using a power law fit on the data between 0.6 K and T$_N$. Integrating $C_{mag}\;dT/T$, from 0 to the current temperature, shifting the values above $T_N$ by the amount of the first-order entropy jump $\Delta S = 0.55$ J/(K mol) (see section \ref{SpecHeat}), one obtains the temperature variation of $S_{mag}(T)$ represented by the cross symbols in Fig. \ref{Entropy}. This temperature dependence shows that the paramagnetic entropy plateau, expected for the cubic $\Gamma_8$ quadruplet CEF ground state, is recovered at temperatures in excess of 30 K. The uncorrected data (empty squares in Fig. \ref{Entropy}) cannot reach the quadruplet value in the graph temperature range. Note that, in agreement with the literature\cite{Lee1972, Fujita1980, PEYSSON1986}, the uncorrected value just above T$_N$ is very close to the entropy of a doublet.\newline
Although the corrected curve asymptotically approaches the $R \ln4$ line in Fig. \ref{Entropy}, one should remember that this variation is subjected to a number of uncertainties. In addition to the discrete integration errors, the jump in entropy at T$_N$ is defined at no better than 10 \%, which represents a potential shift by 0.1 J/(K mol). Moreover, the neutron scattering value $\hbar\omega_0 = 12.7 \text{ meV}$, of limited precision, is directly responsible for the almost zero slope of $S_{mag}(T)$ above 30 K. Nevertheless, it is here shown that the experiments are consistent with a quadruplet CEF ground state and that the recovery of the fourfold degeneracy entropy is progressive: at T$_Q$, the entropy is close to 8 J/(K mol) and it takes more than 30 K to retrieve the missing 30 \% of the quadruplet entropy. 

\section{Entropy of a cage-split quadruplet}
The experiments show that, in the paramagnetic range, the magnetic entropy value just above T$_Q$ is much lower than expected for a quadruplet. It is about $R \ln 2.6$ J/(K mol) against $R \ln4$ for an effective fourfold degeneracy. This means that the ground-state degeneracy is already largely reduced before any ordering process. Such a premature reduction of the entropy is usually ascribed to pair correlations that precede the actual, long range, ordering. However, in the CeB$_6$ case, the ordering temperatures T$_Q$ and T$_N$ are one order of magnitude smaller than the thermal amplitude of the paramagnetic entropy variation. This attests to the weakness of the pair-couplings that drive the ordering, with respect to the energy scale relevant to the entropy variation. This scale actually fits with another one, highlighted in the Raman and neutron scattering investigation of CeB$_6$ crystal field scheme\cite{Zirngiebl1984}. At low temperature, the authors observed an increase in the energy transfer between the $\Gamma _8$ ground state and the excited $\Gamma _7$ level. They interpreted this as resulting from a split $\Gamma _8$ ground state over a 30 K interval, which agrees with the temperature range of the entropy variation. As there is no evidence of a static lattice distortion, the average symmetry of the Ce site remaining cubic, they evoked a possibly dynamic symmetry lowering. This is precisely what can be expected from the large amplitude movement of the rare-earth inside its boron cage, if one considers its crystal field consequences\cite{Amara2019}. In the following, we apply the general considerations of Ref. \onlinecite{Amara2019} to the particular case of CeB$_6$, with the intent of a quantitative description of the thermodynamic anomalies in the paramagnetic range.

\subsection{Cage crystal field}
In case of an offset position of the rare-earth by $\bm{r}$, in addition to the central cubic term $\mathcal{H}_{0}$, the crystal field hamiltonian for the Ce$^{3+}$ ion has to include a correction $\mathcal{H}_{d}(\bm{r})$. As the dynamics of the massive rare-earth is negligible with respect to that of the 4$f$ electron, the here considered correction is static and amounts to a coupling between the 4$f$ quadrupoles and the deviation $\bm{r}$ of the rare-earth from the cage center:
\begin{widetext}
\begin{equation}
\label{HJT}
\mathcal{H}_{d}(\bm{r})=-D^{\gamma} [(3 z^2-r^2) \mathcal{O}_{2}^{0} + 3 (x^2-y^2) \mathcal{O}_{2}^{2}]
-D^{\varepsilon} [ x y \;\mathcal{P}_{xy} + y z \;\mathcal{P}_{yz} + z x \;\mathcal{P}_{zx}]
\end{equation}   
\end{widetext}

where $x$, $y$ and $z$ are the components, along the cubic axes, of the displacement $\bm{r}$ of the rare-earth nucleus from the center of the cage. $\{\mathcal{O}_{2}^{0}$, $\mathcal{O}_{2}^{2}\}$ and $\{\mathcal{P}_{xy}$, $\mathcal{P}_{yz}$, $\mathcal{P}_{zx}\}$ are the quadrupolar operators transforming, respectively, as the $\gamma$ ($\Gamma_3$) and $\varepsilon$ ($\Gamma_5$) cubic representations. In the $J=5/2$ manifold of the Ce$^{3+}$ ion, they are conveniently written in terms of Stevens equivalents \cite{Stevens1952}. $D^{\gamma}$ and $D^{\varepsilon}$ are constants that, within a representation, define the magnitude of the coupling of the 4$f$ quadrupoles with the environment.\newline
At a given position $\bm{r}$ inside the cage, diagonalization of the local crystal field hamiltonian $\mathcal{H}(\bm{r})=\mathcal{H}_{0}+\mathcal{H}_{d}(\bm{r})$ yields the local crystal field scheme and the eigenstates with their composition in terms of $|J, m_J \rangle$ vectors. Inside the cage, the electrostatic energy of the rare-earth ion thus acquires a spatial dependence that contributes to the effective potential well in which it moves. This is the mechanism of the centrifugal Jahn-Teller effect, as described in Ref. \onlinecite{Amara2019}. However, this crystal field contribution is a small correction to the main, non-magnetic, potential term responsible for an energy separation of $\hbar \omega_0 /k_B \approx 150\;$K between the vibration levels. In the following, the cage potential well is therefore considered unaltered by the CEF correction and, consequently, temperature independent. The individual vibrational states of a rare-earth inside a cage are then also temperature independent. As phonons become thermally excited, these stationary states get mixed. However, at temperatures below 50 K, only low frequency phonons get populated: the rare-earth distribution inside the cage remains essentially that of the vibrational ground state, the slight equilibrium shifts induced by phonons resulting, on average, in a tiny widening of the distribution.
In the following calculations, that apply for low temperatures, the only considered rare-earth distribution is that of the unperturbed cage vibrational ground state, thus neglecting : \newline 
- the Jahn-Teller correction,\newline
- the interference of excited vibrational states.

\subsection{Vibrational ground state distribution}
With a position dependent crystal field scheme, describing the properties of the paramagnetic state requires knowledge of the rare-earth distribution inside the cage. There is no direct and precise experimental determination of this distribution: spectroscopic or diffraction approaches all require some modeling or intrinsically lack precision. As regards the spectroscopy, for a given energy separation between the lowest vibration levels, the simplest, cubic or higher symmetry, potential wells all result in similar distributions for the singlet ground state. For consistency with the phonons’ dispersion analysis, we will stick with the harmonic approximation. In this hypothesis, the gaussian wave function of the singlet ground state is entirely defined by the frequency $\omega_0$ and the mass $m$ of the rare-earth. The associated distribution reads as :
\begin{equation}
\label{Distfond}
\rho_{0} (\bm{r}) =  \left( m \omega_{0} \over {\hbar \pi} \right)^{3/2} e^{-\frac{m \omega_{0} r^{2} }{\hbar} }
\end{equation}
for which the full width at half maximum is $w_{HM}=2\sqrt{{\hbar \ln2}/{m \omega_{0} }} $. Using the INS value $\hbar \omega_{0} = 12.7$ meV and the mass of Ce, one obtains $w_{HM} \simeq 0.08 $ {\AA} for CeB$_6$. This amounts to the order of magnitude of an average 4$f$ shell radius and cannot, in relative terms, be neglected\cite{Amara2019}. 
Then, at temperature $T$, the paramagnetic value $\widetilde{\mathcal{A}}(T)$ of an observable $\mathcal{A}$ of the rare-earth ion is the cage average: 
\begin{equation}
\label{average}
\widetilde{ \mathcal{A}}(T) =  \iiint_{V} \; \rho_{0} (\bm{r}) \; \langle \mathcal{A}(\bm{r})\rangle_T \; d^{3}r 
\end{equation}
where $\langle \mathcal{A}(\bm{r})\rangle_T$ is the statistical value of $\mathcal{A}$ at position $\bm{r}$. This value can be considered as statistically defined, at least as a time average. Alternatively, even if $\mathcal{A}$ cannot be defined at $\bm{r}$ for a "reasonable" duration, there are as many instances of the $\bm{r}$ position as there are identical cages in the paramagnetic crystal: $\widetilde{ \mathcal{A}}(T)$ is a macroscopic variable, for which Eq. (\ref{average}) is actually a convenient local definition. In principle, the volume $V$ of the integral should be infinite, but can be restricted to a volume covering the cage extension without significant incidence on the $\widetilde{ \mathcal{A}}(T)$ value. This requires to adapt the normalization of $\rho_{0} (\bm{r})$ to the retained volume. Here, the chosen $V$ volume is a cube of edge $a=0.3$ {\AA}, which is more than three times the FWHM of the distribution. In a numerical implementation of the sum of Eq. (\ref{average}), one can take advantage, as in the calculation of the phonons specific heat, of the cubic symmetry. This is achieved by restricting to samples in $V$ that belong to a representative volume, analog in direct space to the tetrahedron of Fig. \ref{FirstZB}. The numerical results that follow are obtained by considering 286 independent samples (i.e. positions where the hamiltonian is diagonalized and local observables produced), that represent a total of 9261 samples in $V$.

\subsection{Calculation of the thermodynamic functions}
At each sample position $\bm{r_s}$ inside the volume $V$, the total crystal field hamiltonian $\mathcal{H}_{CEF}(\bm{r_s})=\mathcal{H}_{0}+\mathcal{H}_{d}(\bm{r_s})$ has to be numerically diagonalized. Then, at a given temperature T, from the local partition function $Z(T,\bm{r_s})$, the local internal and free energies are derived, as well as the associated entropy. The cage averages are then computed, according to the above described method. This procedure requires a value for the fourth order $B_4$ CEF parameter that defines the cubic $\mathcal{H}_{0}$ crystal field hamiltonian\cite{MorinSchmitt1990} and two other values, $D^{\gamma}$ and $D^{\varepsilon}$, for the displacement-quadrupoles coupling constants of $\mathcal{H}_{d}$. The value $B_4 = -1.47$ K is known from the Raman and INS investigation\cite{Zirngiebl1984}, the negative sign yielding a $\Gamma_8$ ground state. There is no such experimental determination for the $D^{\gamma}$ and $D^{\varepsilon}$ parameters, here introduced for describing the cage-splitting of the $\Gamma_8$ level. One predicted consequence of this splitting in an anomalous variation of the magnetic entropy in the paramagnetic range\cite{Amara2019}, as the one observed in CeB$_6$. Here, it is assumed that this anomaly is entirely due to a cage split $\Gamma_8$. Within this assumption, the values for $D^{\gamma}$ and $D^{\varepsilon}$ are those that best describe the observed temperature variation of CeB$_6$ entropy in the paramagnetic range. From guessed initial values for $D^{\gamma}$ and $D^{\varepsilon}$, a least squares approach was used for optimizing the description of the entropy data in the 4-20 K temperature range (see Fig. \ref{Entropy}). In order to speed up the optimization process, only four experimental points, at T = 4, 10, 12.5 and 20 K, were used for defining the $S_Q (D^{\gamma}, D^{\varepsilon})$ sum of squared differences to be minimized. This search points to four sets of parameters, reported in Table \ref{DgDeParams}, along with their associated $S_Q$ values.

\begin{table}[h]
\centering
\begin{tabular}{c|c|c|c|}

 Set & $D^{\gamma}$ (K/\AA$^2$) & $D^{\varepsilon}$ (K/\AA$^2$) & $S_Q$  (J$^2$/(K$^2$ mol$^2$))  \\
 \hline
 $++$ & +104.1 & +3536  & 3.95x10$^{-4}$\\
\hline
 $+-$ & +75.6 & -4878  &  2x10$^{-4}$\\
\hline
 $-+$ & -107.8 & +3610  & 2.87x10$^{-4}$\\
\hline
$--$ & -98.7 & -4381  &  1.3x10$^{-4}$\\
\hline
\end{tabular}
\caption{\label{DgDeParams} Table of the retained sets of $D^{\gamma}$ and $D^{\varepsilon}$ values that best describe the paramagnetic variation of CeB$_6$ magnetic entropy. The optimization is based on the minimization of $S_Q$, the sum of the squared differences, between experiment and calculation, at four selected temperatures (see text). The sets are named by reference to the respective signs of the parameters.}
\end{table}

\begin{figure}
\includegraphics[width=\columnwidth]{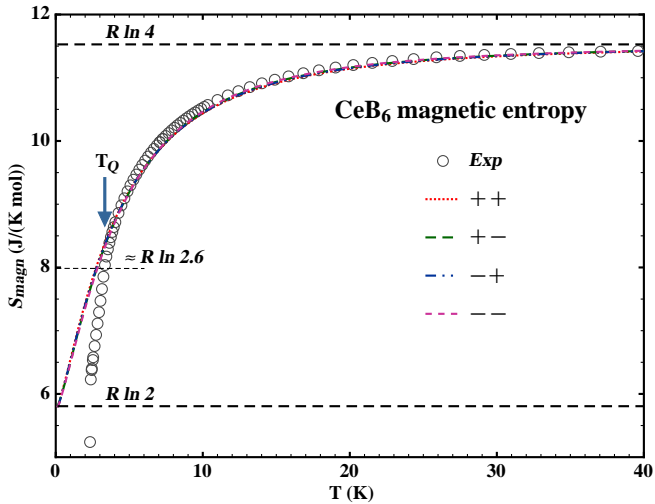}
\caption{\label{EntrMagn} Experimental (circles) and computed variations (lines) of CeB$_6$ magnetic entropy for the  $D^{\gamma}$ and $D^{\varepsilon}$ sets from Table \ref{DgDeParams}. At the graph scale, the computed curves are indistinguishable. The horizontal, dashed, lines give the reference values for a doublet and quadruplet. The shorter dashed line at $\approx R \ln 2.6$ points to the experimental value at T$_Q$.}
\end{figure}

\begin{figure}
\includegraphics[width=\columnwidth]{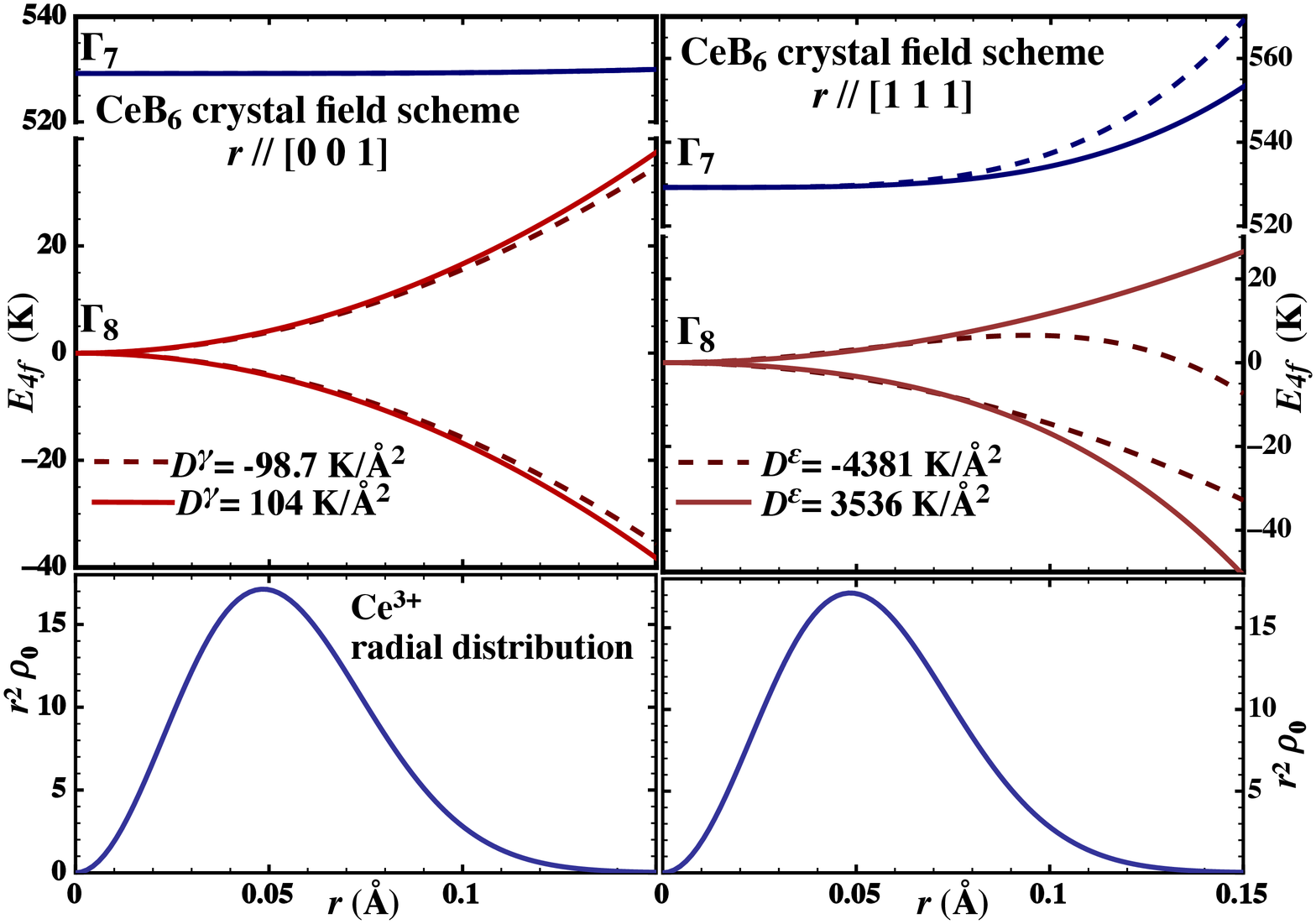}
\caption{\label{Splitting} Cage splitting of the central CEF scheme as result of a displacement $\bm r$ along a fourfold axis (upper left) or threefold axis (upper right). The active quadrupole-displacement coupling constants are respectively $D^{\gamma}$ and $D^{\varepsilon}$. Note that the vertical energy axes are sliced in order to simultaneously display the energies for the $\Gamma_8$ and $\Gamma_7$ levels. The graphs show the effect of $D^{\gamma}$, respectiveley $D^{\varepsilon}$, values with opposite signs, from the sets "$++$" and  "$--$" in Table \ref{DgDeParams}. Lower parts: duplicate radial distributions of Ce$^{3+}$ for the vibrational harmonic ground state, using $\hbar \omega_0 = 12.7$ meV. }
\end{figure}

In Fig. \ref{EntrMagn}, the magnetic entropy variations, computed using the sets of values in Table \ref{DgDeParams}, are compared with our experimental data. Despite an advantage to the sets with negative $D^{\varepsilon}$, in terms of $S_Q$ values, at the graph's scale, the four sets result in indistinguishable lines. They all satisfactorily describe the observed variation of CeB$_6$ entropy in the paramagnetic range. All the zero temperature limits are very close to $R \ln 2$. This corresponds to the expected Kramers minimal degeneracy for Ce$^{3+}$. In the model we use, the fourfold degeneracy of the $\Gamma_8$ level is realized only for a Ce ion at the very center of the cage, which has vanishing weight in the cage averaged values (cf. the distribution in Fig. \ref{Splitting}, lower part). At all locations outside the center, the $\Gamma_8$ quadruplet is split in two doublets. At zero Kelvin, only the lowest local doublet is populated, with corresponding magnetic entropy.

The upper parts of Fig. \ref{Splitting} show the dependencies of the CEF levels as function of the displacement along the two crystallographic directions: fourfold (left) and threefold (right). The two quadrupolar cubic representations are thus separated, $D^{\gamma}$ and $D^{\varepsilon}$ being respectively active for the fourfold axis and threefold axis. Along a fourfold axis, the splitting of the $\Gamma_8$ level has the simplest structure, with a symmetrical energy separation of the two resulting doublets. At the scale of the graph, the upper, $\Gamma_7$, doublet appears unaffected. The splitting scheme along a threefold axis differs on that point, with a substantial interference of the $\Gamma_7$ level at distances greater than 0.05 {\AA} from the center. In particular, for a negative $D^{\varepsilon}$, which corresponds to the best agreement with the entropy results, the mixing between the $\Gamma_7$ and $\Gamma_8$ states induces a strong "repulsion" of the split levels. At distances $r$ above 0.1 {\AA}, the doublets originating from the $\Gamma_8$ plunges to lower energies, while the $\Gamma_7$ goes up. However, the actual consequences of this modified CEF scheme are mitigated by the distribution of the Ce$^{3+}$ ion (see Fig. \ref{Splitting}, lower part), which limits the contribution of distances above 0.1 {\AA}. The influence of the cage split CEF scheme mainly results from the splitting of the $\Gamma_8$ level for distances $r$ around 0.05 {\AA}, where, for the retained $D^{\gamma}$ and $D^{\varepsilon}$ values, the splitting is of similar amplitude (about 8 K) for these two displacement directions.

\begin{figure}
\includegraphics[width=\columnwidth]{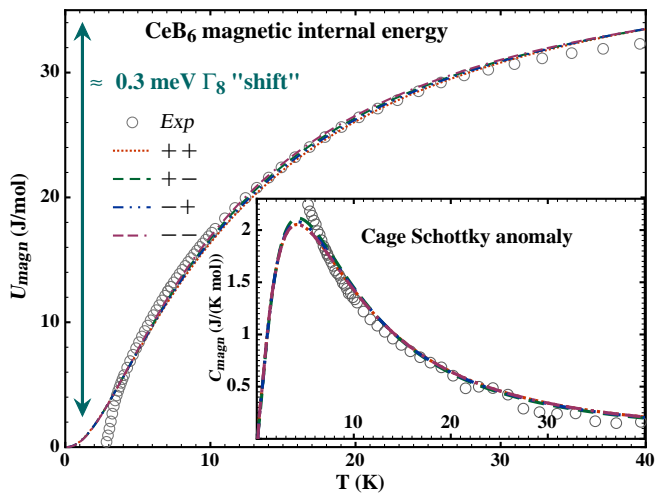}
\caption{\label{EnIntCv} Experimental (circles) and computed (lines) temperature variations of the magnetic internal energy, for the sets of $D^{\gamma}$ and $D^{\varepsilon}$ values in Table \ref{DgDeParams}. The chosen origin for the computed curves is zero energy at 0 K. The experimental data are, accordingly, vertically shifted. This energy variation is indicative of an overall $\approx $ 0.3 meV "shift" in the average energy of the $\Gamma_8$ level. The inset shows the corresponding specific heat curves. The computed ones are obtained from numerical derivation and show the cage Schottky anomaly.}
\end{figure}

Fig. \ref{EnIntCv} shows the paramagnetic variation of the internal magnetic energy. The four computed curves correspond to the set of parameters of Table \ref{DgDeParams}, that optimally describe the entropy thermal dependence. They are vertically shifted in order to have zero internal energy at 0 K. The experimental curve (circles in Fig. \ref{EnIntCv}), obtained by numerical integration of the magnetic part of the specific heat, is also shifted in order to be superimposed with the computed variations. Without surprise, the theoretical descriptions show an agreement of similar quality as observed for the entropy. The well separated $\Gamma_7$ level having negligible influence in the considered temperature range, the dependence of the computed internal energy essentially reflects the Boltzmann population of CEF states originating from the $\Gamma_8$ ground state. The energy curves actually follow the shift of the barycenter of the split $\Gamma_8$ level. This differs from the unsplit and unpopulated $\Gamma_7$ level. Its energy barycenter doesn't depend on the temperature (see Fig. \ref{Splitting}, upper part), but only on the cage distribution of Ce$^{3+}$. Consequently, as the temperature is lowered below 50 K, the average energy separation between the $\Gamma_8$ and $\Gamma_7$ levels increases, reflecting the reduction in the average $\Gamma_8$ energy. This is, in effect, what has been observed via Raman and INS scattering \cite{Zirngiebl1984, Loewenhaupt1985}, that reveal an increase in the energy transfer between the two CEF levels. However, a quantitative examination shows a discrepancy between our calculations, that yield a shift of approximately 0.3 meV (see Fig. \ref{EnIntCv}), and the scattering experiments, that point to a larger 1 meV value. It seems that a reduction in the $\Gamma_8$ energy, consistent with the specific heat results, cannot entirely account for the increase in the $\Gamma_8 - \Gamma_7$ separation. This would require a $\Gamma_7$ level pushed to higher energies as the temperature is reduced. As shown in Fig. \ref{Splitting}, right upper frame, this could be consistent with the cage CEF model in case of a low temperature increase of the Ce$^{3+}$ probability of presence along the threefold axes, beyond $r =$ 0.1 {\AA}. This challenges our assumption of a negligible centrifugal Jahn-Teller effect, with almost temperature independent Ce$^{3+}$ distribution. Another possible fault in our analysis is the confusion between the $\Gamma_8 - \Gamma_7$ average energy separation and the peak position in the Raman or INS scattering experiments. This supposes that the scattering probability is independent of the Ce$^{3+}$ position in the cage. In case of enhanced transition probabilities for peripheral positions, the scattering peak would display, in agreement with the upper part of Fig. \ref{Splitting}, a shift larger than expected from the simple average $\Gamma_8 - \Gamma_7$ separation.\newline
The inset of Fig. \ref{EnIntCv} shows the computed magnetic specific heat curves, deduced from the internal energy variations, that display a characteristic Schottky anomaly, as predicted in Ref. \onlinecite{Amara2019}. They are superimposed with the magnetic specific heat experimental data (circles). The cage split $\Gamma_8$ level accounts well for the increase in the specific heat as the temperature falls below 40 K. Contrary to the usual CEF Schottky anomalies, that result from the proximity of discrete CEF levels, the cage Schottky anomaly starts with a steep slope at zero kelvin. This reflects the continuum of available energy levels that result from the split central CEF ground state. In the case of CeB$_6$, the orderings at T$_Q$ and T$_N$ dominate the low temperature part of the specific heat: no Schottky peak is visible in the experimental data, but the excess of specific heat above T$_Q$ could correspond with the flank of the expected anomaly.

\section{Summary and discussion}
This study follows a long series of investigations of the CeB$_6$ compound. Thanks to new specific heat measurements, the process of the antiferromagnetic transition has been detailed. It is of the first-order kind, with a latent heat of about 1.30 J/mol. Part of the missing paramagnetic entropy of CeB$_6$ is thus recovered at T$_N$. This allows to reconsider the relationship between the so-called antiferroquadrupolar (Phase II) and antiferromagnetic (Phase III) phases. Until then, it was generally considered that the antiferromagnetism was developing over the well-established, non-magnetic, charge organization of Phase II. A first-order transition means that a more drastic transformation can be considered between phases II and III.\newline
Other valuable information can be extracted from the low temperature part of the experimental entropy curve (inset of Fig. \ref{Entropy}). It can be seen that, within phase II, from T$_Q$ to T$_N$, the magnetic entropy decreases from 8 J/(K mol) to 6.2 J/(K mol). This leads, despite the entropy correction of the first-order transition, to an entropy value for phase II at T$_N$ that only slightly exceeds $R \ln 2 = 5.8$ J/(K mol). If phase II were, as reported, non-magnetic, its 0 K extrapolated entropy value should be close to that of a doublet, i.e. $R \ln 2$. This is the Kramers theorem applied to a Ce$^{3+}$ ion. In case of a simple, antiphase, magnetic order, the extrapolated entropy value would be zero, as for phase III. In view of Fig. \ref{Entropy}, it is likely that the 0 K extrapolated value for phase II is much smaller than $R \ln 2$, but still larger than 0. This intermediate value adds to the peculiarity of phase II and puts into question its non-magnetic nature. If the ordering mechanism for phase II were indeed non-magnetic, a 0 K extrapolated entropy value much lower than $R \ln 2$ is, at least, indicative of strong magnetic correlations. Evidence for magnetic correlations within Phase II have been previously obtained from polarized neutron scattering experiments\cite{Plakhty2005}, indicative of short-range magnetic arrangements with $[\frac{1}{2}\;\frac{1}{2}\;\frac{1}{2}]$ wave vector. \newline
The calculation of the phonons contribution to the specific heat of hexaborides, based on a better account of the cage system dispersion curves, yields a satisfactory description for LaB$_6$. It shows that, at temperatures below 50 K, it is the cage oscillator frequency that is the main determinant of the changes across the series. No simple method allows to extrapolate from one element to another: an experimental determination is required. In the case of CeB$_6$, the used value is derived from the phonons' dispersion curves obtained from inelastic neutron scattering.\newline
Thanks to the correction introduced by the first-order magnetic transition and to this improved description of the phonon contribution, an improved experimental determination of CeB$_6$ magnetic entropy has been obtained. The temperature variation of this entropy displays the paramagnetic plateau characteristic of the fourfold degenerate $\Gamma_8$ CEF ground state. This plateau doesn't materialize immediately above the ordering temperature, but for temperatures higher than 30 K. This is ten times higher than the ordering temperature of CeB$_6$ and unlikely to relate to pair correlations.\newline
The temperature scale of this abnormal thermal evolution fits with another peculiarity of CeB$_6$: the CEF excitation, as evidenced by Raman and thermal neutrons scattering, shifts towards higher energies below 20 K. In the cage context, with an orbitally degenerate cubic CEF ground state, these two anomalies can be related to a single mechanism: the dynamical splitting of the CEF ground state as result of the rare-earth movement. An attempt, based on this crystal field mechanism, at describing the entropy anomaly of CeB$_6$ is satisfactory, at the price of the introduction of two parameters describing the CEF change for an offset Ce$^{3+}$. An associated shift with the temperature of the CEF $\Gamma_8 - \Gamma_7$ excitation is predicted, the simplest estimate yielding about one third of the reported value. 
However, as the computed values, confronted with the experimental data, result from Boltzmann and spatial averages, there is some intrinsic indetermination in this CEF description. In the case of CeB$_6$, at least four sets of parameters are consistent with the specific heat data. Additional experimental data may allow to distinguish between them, but this also requires some theoretical effort in order, for instance, to describe experiments under an applied magnetic field.\newline
Another test of this cage crystal field interpretation would be to look for signatures of the associated centrifugal Jahn-Teller effect\cite{Amara2019}. This effect has direct, but moderate, consequences on the system volume and vibration frequency of the rare-earth. In CeB$_6$, as for two other rare-earth hexaborides with non-Kramers CEF ground states, PrB$_6$ and NdB$_6$, X ray diffraction results\cite{Novikov2017} show, in the low temperature paramagnetic range, a thermal expansion anomaly with respect to the LaB$_6$ reference. This might be the expected volume consequence of the Jahn-Teller effect. As regards the changes in the oscillator frequency, a softening of associated phonons in the percent range is expected\cite{Amara2019}. Its detection requires high resolution, infrared or neutron, spectroscopic techniques. Presently, there are no available experimental results fulfilling these requirements.
\newline

\begin{acknowledgments}
The authors would like to thank Dr. N. Y. Shitsevalova and collaborators, from the IPMS in Kiev, Ukraine who prepared the LaB$_6$ single crystals. We are also grateful to Dr. L.-P. Regnault from CEA, INAC in Grenoble, France, for lending us the CeB$_6$ crystals and exchanging views on the CeB$_6$ puzzle.
\end{acknowledgments}

\end{document}